\documentclass[11pt]{article}
\usepackage[letterpaper]{geometry}

\usepackage{graphicx,subfigure}
\graphicspath{{./figures/}} 
\usepackage{amsfonts,amsbsy,amscd,amsgen,amsthm,dsfont,amsmath,amssymb,mathrsfs,mathtools,array}
\usepackage[colorlinks=true,allcolors=blue]{hyperref}
\usepackage{bm,multirow,esvect,cancel}
\theoremstyle{definition}
\usepackage{tabularx}
\usepackage{xcolor,comment,soul}
\usepackage{enumitem}
\usepackage[utf8]{inputenc}
\usepackage[T1]{fontenc}
\usepackage{authblk}
\usepackage[]{algorithm}
\usepackage{algpseudocode} 
\usepackage[font=small,labelfont=bf]{caption}
\usepackage{letltxmacro}
\LetLtxMacro{\originaleqref}{\eqref}
\renewcommand{\eqref}{Eq.~\originaleqref}

\newcommand{\id}{\mathrm d}

\DeclareMathAlphabet\mathbfcal{OMS}{cmsy}{b}{n}

\begin{document}

\title{Stochastic compartmental models of COVID-19 pandemic must have temporally correlated uncertainties}

\author[a]{Konstantinos Mamis}
\author[a]{Mohammad Farazmand\thanks{Corresponding author: farazmand@ncsu.edu}} 
\affil[a]{Department of Mathematics, North Carolina State University,
2311 Stinson Drive, Raleigh, NC 27695-8205, USA}
\date{}

\maketitle

\begin{abstract}
Compartmental models are an important quantitative tool in epidemiology, enabling us to forecast the course of a communicable disease.
However, the model parameters, such as the infectivity rate of the disease, are riddled with uncertainties, which has motivated the development and use of stochastic compartmental models. Here, we first show that a common stochastic model, which treats the uncertainties as white noise, is fundamentally flawed since it erroneously implies that greater parameter uncertainties will lead to the eradication of the disease. Then, we present a principled modeling of the uncertainties based on reasonable assumptions on the contacts of each individual. Using the central limit theorem and Doob's theorem on Gaussian Markov processes, we prove that the correlated Ornstein--Uhlenbeck process is the appropriate tool for modeling uncertainties in the infectivity rate.
We demonstrate our results using a compartmental model of the COVID-19 pandemic and the available US data from the Johns Hopkins University COVID-19 database. In particular, we show that the white noise stochastic model systematically underestimates the severity of the Omicron variant of COVID-19, whereas the Ornstein--Uhlenbeck model correctly forecasts the course of this variant. Moreover, using an SIS model of sexually transmitted disease, we derive an exact closed-form solution for the asymptotic distribution of infected individuals. This analytic result shows that the white noise model underestimates the severity of the pandemic because of unrealistic noise-induced transitions. Our results strongly support the need for temporal correlations in modeling of uncertainties in compartmental models of infectious disease.

\end{abstract}

\section{Introduction}
\begin{figure*}[!t]
	\centering
	\includegraphics[width=\textwidth]{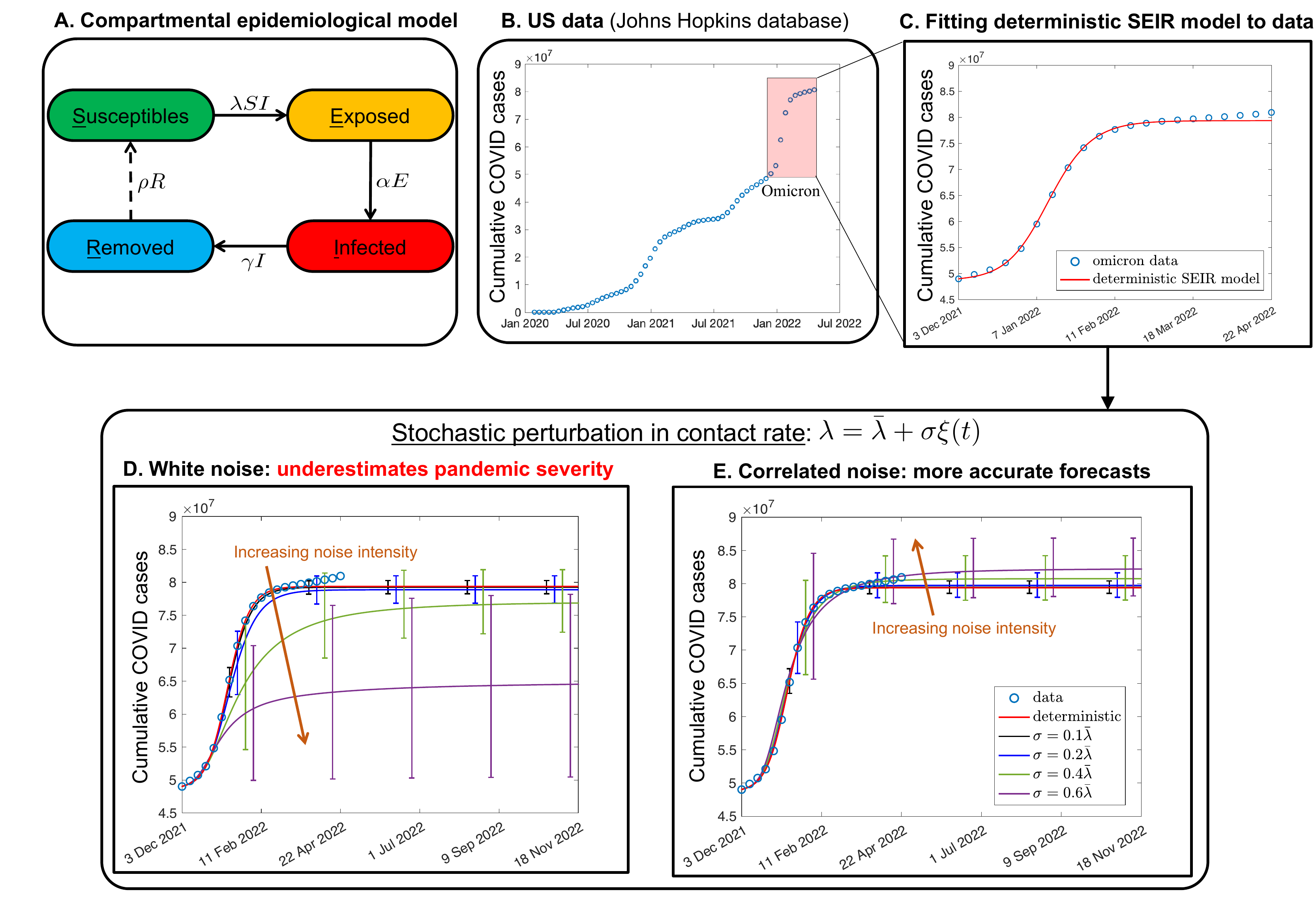}
	\caption{Summary of stochastic modeling and results for the Omicron variant of COVID-19 pandemic in the US. \textbf{A:} SEIRS deterministic compartmental model suitable for airborne diseases, such as COVID-19. The parameters in SEIRS model are: i) the average adequate contact rate $\lambda$ of each individual (also called the infectivity rate), ii) the average incubation rate $\alpha$, which is the inverse of the average incubation (or latency) period, during which the individual has contracted the virus but is not infectious yet,  iii) the average curing rate $\gamma$ which is the inverse of the average time each individual needs to recover, iv) $1/\rho$ is the average period after which immunity wanes off. Links between compartments shown in dashed lines are those omitted in the SEIR model of \eqref{eq:SEIR} we use here.  \textbf{B:} cumulative number of COVID-19 cases in the US. Light red frame highlights the Omicron variant wave (Dec. 2021-Apr. 2022). \textbf{C:} COVID-19 forecast obtained by least-squares fitting of the deterministic SEIR model to the data for Omicron cases. Contact rate $\lambda$ of fitted SEIR model is then stochastically perturbed by noise $\xi(t)$ with intensity $\sigma$, around its mean value $\bar{\lambda}$.  \textbf{D:} Results of white noise stochastic SEIR models for increasing noise intensity $\sigma$, obtained from 50,000 Monte Carlo simulations.  Mean trajectories of COVID cases are shown, equipped with error bars inside which the 50\% of trajectories lie. \textbf{E:} Same as panel D except that $\xi(t)$ is the Ornstein--Uhlenbeck noise with correlation time $\tau=1$ week.}
	\label{fig:1}
\end{figure*}

Quantitative models for forecasting the spread of communicable disease are a valuable tool for policymaking during pandemics, such as the ongoing COVID-19, and endemics, such as gonorrhea. Compartmental models are simple, widely used, quantitative tools for studying the spread of such infectious disease~\cite{Kermack1927,Capasso1978,Hethcote1994,Brauer2008,Capasso2008, Diekmann2021}.
For instance, an appropriate compartmental model for COVID-19 pandemic is the Susceptible-Exposed-Infected-Removed (SEIR) model~\cite{Bertozzi2020,Faranda2020,Faranda2020a,Linka2020,Yakubu2021,Vakil2022,Wu2020,Feng2021,Irons2021}, depicted in Fig. \ref{fig:1}A. As shown in Fig.~\ref{fig:1}C, if the model parameters are chosen properly, the deterministic SEIR model accurately forecasts the spread of the Omicron variant in the US.

However, the model parameters, such as the infectivity rate $\lambda$, the incubation rate $\alpha$ and the curing rate $\gamma$, are a priori unknown. These parameters can be estimated by collecting and averaging data over the population. Such data is inevitably riddled with uncertainties which has motivated the development and use of stochastic compartmental models.

In stochastic compartmental models, one assumes that the model parameters are stochastic processes. The most common approach is to assume that the parameter comprises a constant mean perturbed with white noise~\cite{Gray2011,Dureau2013,Maki2013,Faranda2020a}. For instance, the infectivity or contact rate is often modeled as $\lambda(t) = \bar{\lambda} +\sigma\xi(t)$, where
$\bar\lambda$ is the mean, $\xi(t)$ is the standard white noise, and $\sigma$ is a constant controlling the noise intensity~\cite{Gray2011,Ji2014,Cai2015,Meng2016,Cai2020}. Here, we first show that, although this assumption seems reasonable, it leads to systematic underestimation of the disease spread. For instance, Fig.~\ref{fig:1}D shows the SEIR prediction of the Omicron wave in the US with the infectivity rate $\lambda$ modeled as white noise. The resulting stochastic SEIR model consistently underestimates the true cumulative number of  COVID cases. Making matters worse, as the noise intensity $\sigma$ increases, the white noise stochastic model further underestimates the number of COVID cases.

As we discuss in the Results section, this behavior is not specific to the SEIR model, but is also observed in SIR and SIS models.
This indicates that the stochastic compartmental models with white noise have a fundamental flaw: they imply that greater uncertainty in the model parameters leads to a less severe spread of the disease. This model behavior is certainly dubious since greater ignorance about the disease, e.g., its infectivity rate, does not in reality make a pandemic less severe. The counter-intuitive and unrealistic implications of modeling parameter uncertainties with white noise had also been noted in oncology models, where increasing the noise intensity leads to tumor eradication, a behavior that is not replicated in actual cases~\cite{DOnofrio2008}.

In contrast, as shown in figure~\ref{fig:1}E, modeling uncertainties by correlated noise alleviates such contradictory behavior. Increasing the intensity of correlated noise leads to forecasts that are essentially in line with the deterministic model. But, as one expects, the uncertainties around the stochastic predictions grow as the noise intensity grows.

Here, for the first time, we present a principled modeling of uncertainties in the infectivity rate of epidemiological models. Starting from reasonable assumptions on the contact rate of each individual, and using the central limit theorem and Doob's theorem on Gaussian Markov processes, we show that the only admissible model for such uncertainties is the mean-reverting Ornstein--Uhlenbeck (OU) process. We demonstrate the implications of our results on two examples, SIS model of sexually transmitted disease and SEIR model of COVID-19. For the SIS model, we derive a closed-form solution for the probability distribution of the infected population and show that the OU-based model always has a peak near the deterministic equilibrium.
In contrast, modeling uncertainties with white noise erroneously predicts a less severe disease spread, or even its eradication, as the uncertainties grow.
Similar observations are made for the SEIR model of COVID-19 using Monte Carlo simulations and the Johns Hopkins University database of COVID-19 cases.

We note that different types of noise, other than white noise, have also been used to model uncertainties in biological systems~\cite{Allen2016}. For instance, log-normal noise have been used to model uncertainties in COVID-19 transmission under the presence of superspreaders \cite{Faranda2020a}. 
OU process has already been used in the context of sexually transmitted and bacterial diseases \cite{Wang2018}. However, its rigorous justification from first principles is presented here for the first time.

\section{Stochastic modeling of the average contact rate}
There are several parameters in the compartmental models of infectious disease. For instance, in the SEIR model, these are the infectivity or average contact rate $\lambda$, the incubation rate $\alpha$ and the curing rate $\gamma$. The source of their uncertainty is the underlying dynamical interplay between biological and social factors.  Incubation and curing rates depend mainly on the biology of the virus.  However, the contact rate $\lambda$ depends on both biological factors, e.g., how easily the virus transmits, and social factors, e.g., how many individuals an infected person comes in contact with. As a result, the contact rate is the main source of uncertainty  in epidemiological models. Therefore, for the sake of simplicity, we assume that the average incubation and curing rates are deterministic constants and  that the contact rate $\lambda$ is the only stochastic parameter. This is a common choice in the literature on stochastic compartmental models \cite{Gray2011,Linka2020,Neri2021}.

We denote the total number of adequate contacts of the $n$th individual, up to time $t$, by $C_n(t)$.  A contact is adequate if it is sufficient for the transmission of the virus~\cite{Hethcote1984}. The contact rate of the $n$th individual is then defined as the rate of change of $C_n$, i.e., $\lambda_n = \id C_n/\id t$. Since $C_n(t)$ is not necessarily differentiable, it is more convenient to work with its increments, $\Delta C_n(t)$,  that is the number of contacts of the $n$th individual over the time interval $[t,t+\Delta t]$, where $\Delta t$ is a small time increment.  The contact rate is then approximated by $\lambda_n(t) \simeq \Delta C_n(t)/\Delta t$.

We treat the incremental contacts $\Delta C_n(t)$ of each individual as a stochastic process satisfying the following assumptions:
\begin{enumerate}[label = (\arabic*)]
\item There is no dependence between the incremental contacts of different individuals. In other words, we assume  $\left\{\Delta C_n(t)\right\}_{n=1}^N$ is a collection of $N$ independent random processes. 
\item The number of contacts each individual makes in $[t,t+\Delta t]$ does not depend on the history of their contacts in previous time instances; thus $\Delta C_n(t)$ is a Markov process, where $N$ is the population size.
\item The mean and covariance of the incremental contacts satisfy 
\begin{subequations}\label{eq:Cn_mc}
\begin{equation}\label{eq:mean}
	\mathbb{E}\left[\Delta C_n(t)\right]=\mu\Delta t,
\end{equation}
\begin{equation}\label{eq:covariance}
	\mathrm{Cov}\left[\Delta C_n(t),\Delta C_n(s)\right]=\frac{\sigma_0^2}{2\tau}\exp\left(-\frac{|t-s|}{\tau}\right)\Delta t^2,
\end{equation}
\end{subequations}
where $[t,t+\Delta t]$ and $[s,s+\Delta t]$ are disjoint time intervals and $\tau$ is the correlation time.  
\end{enumerate}

\eqref{eq:Cn_mc} models the assumption that the incremental contacts of every individual are proportional to the time interval $\Delta t$. \eqref{eq:mean}, where $\mu$ is a constant, models the assumption of no seasonal variability in contacts. The exponential factor in \eqref{eq:covariance} constitutes the simplest model for temporally correlated incremental contacts with finite correlation time $\tau$. Note that, although the incremental constants $\Delta C_n(t)$ have the same mean and covariance, we do not assume that they are identically distributed.

Assumptions  (1)-(3) determine the properties of the contact rate $\lambda_n(t)$.
In particular, $\lambda_n(t)$ inherits the Markovian property of $\Delta C_n(t)$. Also,  the mean value and covariance of $\lambda_n(t)$ are determined via Eq.~(\ref{eq:Cn_mc}),
\begin{subequations}
\begin{equation}\label{eq:mean2}
\mathbb{E}\left[\lambda_n(t)\right]=\mu,
\end{equation}
\begin{equation}\label{eq:covariance2}
\mathrm{Cov}\left[\lambda_n(t),\lambda_n(s)\right]=\frac{\sigma_0^2}{2\tau}\exp\left(-\frac{|t-s|}{\tau}\right).
\end{equation}
\end{subequations} 
The exponentially decaying correlation of \eqref{eq:covariance2} for the contact rate of each individual was also assumed in the agent-based model of Ariel and Louzun~\cite{Ariel2021}.
 
Having defined the contact rate $\lambda_n(t)$ for every individual, we now introduce the contact rate averaged over the whole population as $\lambda(t)=(1/N)\sum_{n=1}^N\lambda_n(t)$. 

Now,  we consider $K$ time instances $t_1<\cdots<t_k<\cdots<t_K$ and define the vectors $\bm{\lambda}_n=[\lambda_n(t_1),\cdots,\lambda_n(t_k),\cdots,\lambda_n(t_K)]^T$, which samples the stochastic process $\lambda_n(t)$ at the said time instances. The random vector $\bm{\lambda}_n$ has mean value $\bm{\mu}$ with $\mu_k=\mu$,  and covariance matrix $\bm{\Sigma}$ with $\Sigma_{k\ell}=\mathrm{Cov}\left[\lambda_n(t_k),\lambda_n(t_\ell)\right]$, $k,\ell=1,\ldots K$. For the average vector $\bm{\lambda}=(1/N)\sum_{n=1}^N\pmb\lambda_n$, the multidimensional central limit theorem \cite{Vaart1998} implies 
\begin{equation}
\sqrt{N}(\bm{\lambda}-\bm{\mu})\sim\mathcal{N}_K(0,\bm{\Sigma}),
\end{equation}
where $N$ is the total population, and $\mathcal{N}_K$ is the $K$-variate normal distribution.  
Note that $\bm\lambda$ can be viewed as the discrete samples of the average contact rate $\lambda(t)$, so that 
$\bm{\lambda}=[\lambda(t_1),\cdots,\lambda(t_k),\cdots,\lambda(t_K)]^T$.

Thus, the stochastic process $\lambda(t)$ of the average contact rate has the following properties:
\begin{enumerate}[label=(\roman*)]
\item It is temporally homogeneous, since for every sequence $t_1<\cdots<t_k<\cdots<t_K$, all the $K$-variate joint distributions of $\lambda(t_1),\cdots,\lambda(t_K)$ are invariant to translations in time.
\item It is a Markov process, since it is the average of independent Markov processes \cite{Stoyanov1987,Rozhkov2010}.
\item For every pair of two time instances, $t$ and $s$, its values $\lambda(t)$ and $\lambda(s)$ follow a bivariate Gaussian distribution. 
\end{enumerate}
Doob's theorem on Gaussian Markov processes~\cite{Doob1942} implies that the only two possible stochastic processes with properties (i)-(iii) are the uncorrelated white noise and the correlated Ornstein--Uhlenbeck noise. White noise is also ruled out since we have already proven by the central limit theorem that $\lambda(t)$ and $\lambda(s)$ are not independent. Therefore, the only choice for $\lambda(t)$  is to be an Ornstein--Uhlenbeck process, with mean value
\begin{equation}\label{eq:mean3}
\mathbb{E}\left[\lambda(t)\right]=\mu.
\end{equation}
and autocovariance
\begin{equation}\label{eq:covariance3}
\mathrm{Cov}\left[\lambda(t),\lambda(s)\right]=\frac{\sigma^2}{2\tau}\exp\left(-\frac{|t-s|}{\tau}\right),
\end{equation}
with $\sigma=\sigma_0/\sqrt{N}$.  

It is well-known that the Ornstein--Uhlenbeck process, with mean (\ref{eq:mean3}) and stationary autocovariance (\ref{eq:covariance3}), is generated by the stochastic differential equation,
\begin{equation}
\id \lambda = \frac{1}{\tau}(\mu - \lambda)\id t +\frac{ \sigma}{\tau} \id W,
\end{equation}
where $W(t)$ is the standard Wiener process~\cite{Hanggi1995}. We note that, as the correlation time $\tau$ tends to zero, the uncorrelated white noise is retrieved \cite[Sec. 6.6]{Toral2014}. In the Results section, we use this limiting relationship to compare results from stochastic compartmental models under white and Ornstein--Uhlenbeck noise.

For notational simplicity, we write $\lambda(t)=\bar{\lambda}+\sigma\xi(t)$,  where $\bar{\lambda}=\mu$ is its mean value, $\sigma$ is its noise intensity, and $\xi(t)$ is a standard Ornstein--Uhlenbeck process with zero mean and autocovariance given by \eqref{eq:covariance3} with $\sigma=1$.  

In the course of our work, we became aware of a recent study~\cite{Jing2022} which also invokes the multidimensional central limit theorem, in the context of SIS equation on networks, to model uncertainties as white noise. 
However, as we argued above and show in the Results section, the correct model for uncertainties is the correlated OU process.

\section{Results}
\subsection{Stochastic SIS models}
Before discussing the COVID-19 data, we examine the simpler SIS model where we derive the probability distribution of the 
infected individuals in closed form. This allows us to highlight the differences between the white noise and OU noise with no ambiguity.
The SIS model is governed by the equations
\begin{subequations}\label{eq:SIS}
	\begin{equation}\label{eq:S}
		\frac{\id S}{\id t}=-\frac{\lambda}{N}SI+\gamma I,
	\end{equation}
	\begin{equation}\label{eq:I}
		\frac{\id I}{\id t}=\frac{\lambda}{N}SI-\gamma I,
	\end{equation}
\end{subequations}
with the susceptible (S) and infected (I) compartments where $N=S+I$ is the total population. 
We assume that the curing rate $\gamma$ is a deterministic constant and the contact rate $\lambda(t)=\bar\lambda+\sigma\xi(t)$ is a stochastic process.
We consider two types of noise $\xi(t)$: white noise and the standard OU noise. It is customary to express the results in terms of the infected fraction $X= I/N$. Note that since the total population $N$ is conserved, we have $S= N(1-X)$.

For stochastic SIS models, we are able to obtain the asymptotic distribution of the infected fraction $X$ of the population, in closed form (see Appendix \ref{sec:stochastic_models}).  As seen in Fig. \ref{fig:3}, the theoretical asymptotic distributions are always in excellent agreement with Monte Carlo simulations. 

\begin{figure*}[!t]
	\centering
	\includegraphics[width=\textwidth]{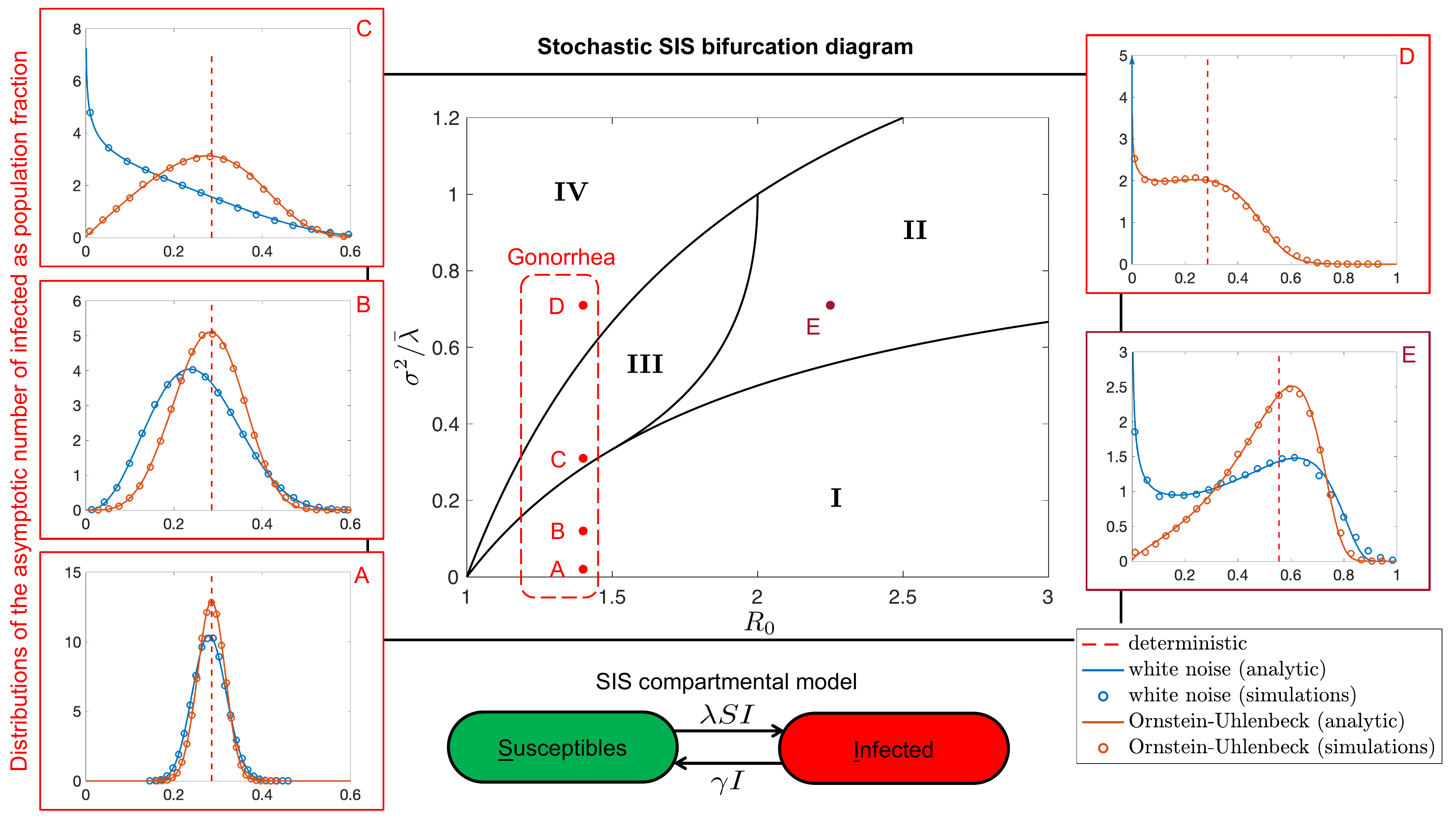}
	\caption{\textit{(central figure)} Bifurcation diagram for the SIS model under white noise perturbation of its contact rate, depending on two dimensionless parameters, the basic reproduction number $R_0=\bar{\lambda}/\gamma$ and the relative noise variance $\sigma^2/\bar{\lambda}$. The four regions shown depict different shapes for the distribution of the asymptotic number of infected $X$ as a fraction of the population. \textbf{I:} Unimodal with mode at a non-zero $X$.  \textbf{II:} Bimodal with one mode at $X=0$. \textbf{III:} Unimodal with mode at $X=0$. \textbf{IV:} Delta function at $X=0$. \textit{(peripheral figures A-D)} Distributions for the asymptotic infected fraction for gonorrhea ($R_0=\bar{\lambda}/\gamma=1.4$, $\gamma=1/20$ days$^{-1}$ \cite{Hethcote1984}) for increasing noise intensity: $\sigma=0.2\bar{\lambda}$ for panel A, $\sigma=0.5\bar{\lambda}$ for B, $\sigma=0.8\bar{\lambda}$ for C, $\sigma=1.2\bar{\lambda}$ for D.  In each figure, the stable equilibrium of the original deterministic model is marked by a red dashed line.  The points corresponding to panels A-D cases are also marked on the central figure. The additional case E, not corresponding to a disease, is also shown as a representative of region II of the bifurcation diagram. In peripheral figures,  we also depict the distribution under Ornstein--Uhlenbeck noise with correlation time half the characteristic time scale of the system (see Materials and Methods).  Open circles correspond to distributions obtained from 50,000 samples of direct Monte Carlo simulations, while solid lines mark the exact distributions. In panel D, the distribution corresponding to white noise perturbation is a delta function at $X=0$, and thus it is depicted as a vertical arrow.}\label{fig:3}
\end{figure*}

The asymptotic distribution depends on two dimensionless parameters (see Appendix \ref{sec:stochastic_models}), the deterministic basic reproduction number $R_0=\bar{\lambda}/\gamma$ and the relative variance $\sigma^2/\bar{\lambda}$ of the noise.  Thus, we are able to produce the bifurcation diagram shown in Fig. \ref{fig:3},  where the possible asymptotic forms of the distribution are depicted.  For $R_0>1$, the deterministic SIS model ($\sigma=0$),  has one stable equilibrium at $X=(\bar{\lambda}-\gamma)/\bar{\lambda}$, and one unstable equilibrium at $X=0$. However, the stochastic SIS model with white noise exhibits a richer asymptotic behavior that includes parameter regions of bistability and regions where $X=0$ is stable.

We first focus on $R_0=1.4$ which is the relevant basic reproduction number for gonorrhea \cite{Hethcote1984}.  As shown in Fig. \ref{fig:3}, by increasing white  noise intensity, the infected fraction distribution undergoes a transition. For low noise intensity, the distribution exhibits one mode, close to the stable deterministic equilibrium (see Fig. \ref{fig:3}A). As the noise intensity increases, the distribution mode shifts towards zero (see Fig.~\ref{fig:3} B, C). 
Increase the white noise intensity further results in the eradication of the disease from the population, since the distribution of the infected fraction $X$ becomes a delta function at zero (see Fig.~\ref{fig:3}D). We emphasize that this is an unrealistic model behavior since it implies that greater ignorance about the contact rate $\lambda$ will lead to disease eradication. 

In Figs.~\ref{fig:3}A-D, we also plot the distribution obtained under an Ornstein--Uhlenbeck perturbation of the contact rate in the SIS model.  As correlation time of the Ornstein--Uhlenbeck noise, we choose $\tau=0.5(\bar{\lambda}-\gamma)^{-1}$, where $(\bar{\lambda}-\gamma)^{-1}$ is the characteristic time scale of the deterministic SIS model (see Materials and Methods).

We observe that, compared to white noise perturbation, correlation in noise has the effect of suppressing transition away from the deterministic equilibrium.  For noise levels as high as $80\%$, the mode of the distribution remains close to the stable deterministic equilibrium. For very high noise levels (such as $\sigma=1.2\bar{\lambda}$ of Fig. \ref{fig:3}D) the distribution under Ornstein--Uhlenbeck perturbation also exhibits an additional mode at 0. However, most of distribution mass is still located around the deterministic equilibrium, whereas, for the same noise levels, the SIS model with white noise perturbation predicts the eradication of the disease from the population.

\subsection{Omicron variant in the US}
Next, we study the Omicron variant of COVID-19 in the US, using the SEIR model, 
\begin{subequations}\label{eq:SEIR}
	\begin{equation}
	\frac{\id S}{\id t}=-\frac{\lambda}{N}SI
	\end{equation}
	\begin{equation}
	\frac{\id E}{\id t}=\frac{\lambda}{N}SI-\alpha E
	\end{equation}
	\begin{equation}
	\frac{\id I}{\id t}=\alpha E-\gamma I
	\end{equation}
	\begin{equation}\label{eq:removed}
	\frac{\id R}{\id t}=\gamma I,
	\end{equation}
\end{subequations}
with the susceptible (S), exposed (E), infected (I), and removed (R) compartments. It shares the same set of parameters with the SIS model as well as the additional parameter $\alpha$ denoting the average incubation rate. We note that more complex models which include a vaccinated compartment, e.g., the Susceptible-Vaccinated-Exposed-Infected-Removed (SVEIR) model \cite{Ghostine2021}, can also be used. 
Following earlier studies~\cite{Bertozzi2020,Faranda2020,Faranda2020a,Linka2020,Yakubu2021,Vakil2022,Wu2020,Feng2021,Irons2021}, here we treat individuals that are effectively vaccinated as part of the removed compartment and therefore use the simpler SEIR model. 

Our main COVID-19 results are shown in Fig.~\ref{fig:1}. First, we determine the parameters of the deterministic SEIR model by fitting its trajectory to the data from the cumulative COVID-19 cases in the US \cite{JohnsHopkins} for the period of December 2021-April 2022, during the Omicron wave (see Fig.~\ref{fig:1}B,C).  The fitting procedure is described in Materials and Methods.  Then we consider the stochastic contact rate $\lambda(t) = \bar\lambda+\sigma\xi(t)$ which is perturbed around its deterministic value $\bar\lambda=1.54$ days$^{-1}$. For the perturbation $\xi(t)$, we consider both white and Ornstein--Uhlenbeck processes, and perform Monte Carlo simulations of the respective stochastic SEIR models (see Fig.~\ref{fig:1}D,E). The noise intensity is varied between 10\% and 60\% of the average contact rate $\bar{\lambda}$.  For the Ornstein--Uhlenbeck noise, we also have to choose the value of the correlation time $\tau$.  Despite the large number of studies, precise data on the temporal correlation of contact rates is scarce \cite{Tkachenko2021}.  For the results shown in Fig.~\ref{fig:1}E, we choose one week as the correlation time, motivated by the weekly pattern in human activity~\cite{Goh2008,Jiang2012,Kivela2012,Saramaki2015}. The results are similar for shorter and longer correlation times (see Appendix \ref{sec:SEIR}).

\begin{figure*}
	\centering
	\includegraphics[width=\textwidth]{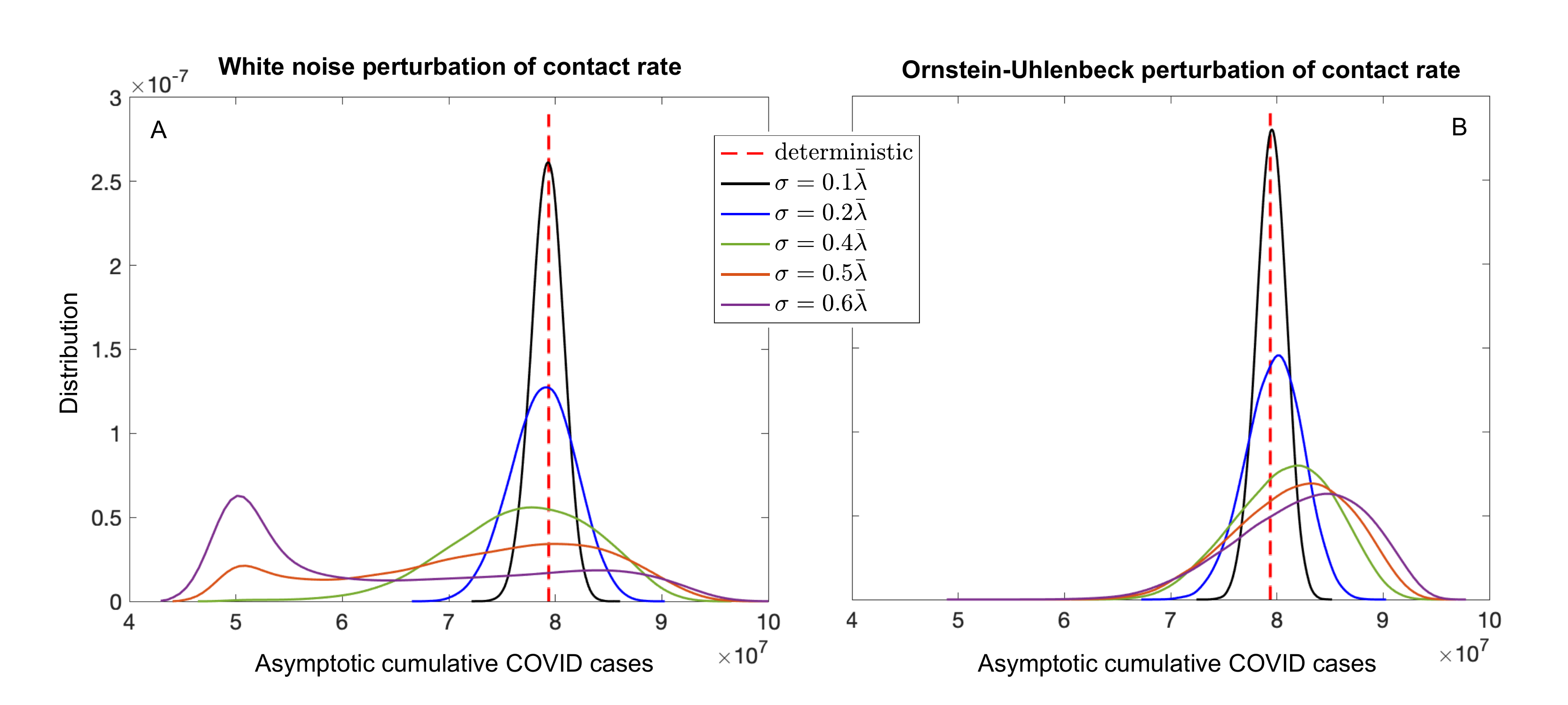}
	\caption{Distributions of the cumulative COVID cases in the US, as determined from SEIR models with stochastically perturbed contact rate.  Model parameters are the same as in Fig. \ref{fig:1}. Stable equilibrium point of the deterministic SEIR is shown as red dashed line. \textbf{A:} SEIR with contact rate perturbed by white noise. \textbf{B:} SEIR with contact rate perturbed by Ornstein--Uhlenbeck noise with correlation time $\tau=1$ week.}\label{fig:2}
\end{figure*}

As shown in Fig.~\ref{fig:1}D, increasing the intensity of both white and OU noise results in an expected increase in variance of the forecasted omicron cases. However, we observe that, in the white noise case, increasing the noise intensity results in the mean to underestimate the asymptotic number of cases, predicting significantly less spread of the disease in the population. For instance, when the noise intensity is at 60\%, the forecasted cumulative number of COVID cases is underestimated by approximately 20\%.  This result is unrealistic, since higher uncertainty in the contact rate does not in reality result in a less severe pandemic.  

On the other hand, the results in Fig. \ref{fig:1}E for the correlated Ornstein--Uhlenbeck noise are reliable. The mean trajectories stay close to the deterministic prediction and the data. As the noise intensity increases, the predictions attain slightly higher mean values of the asymptotic number of cases, compared to the value predicted by the deterministic model. For instance, at 60\% noise intensity, the forecasted cumulative number of COVID cases increases by approximately 4\%.

The mechanism by which SEIR model under white noise underestimates the size of COVID-19 pandemic is better understood by inspecting the noise-induced qualitative changes in distribution of the asymptotic number of cases (see Fig. \ref{fig:2}). In absence of a closed-form solution, we use Monte Carlo simulations to determine the distribution corresponding to the stochastic SEIR model. 

As shown in Fig. \ref{fig:2}, for small noise intensities, the distributions of both white and OU stochastic models are unimodal and narrow, with their modes approximately coinciding with the deterministic equilibrium, marked by a dashed line. The distribution under Ornstein--Uhlenbeck noise is slightly less diffusive than the one under the white noise perturbation, which is consistent with the \textit{sharpening effect of correlated noise} \cite{Hanggi1995,Mamis2021}.  As expected, increase in noise intensity results in distributions that are more diffusive, and in distribution modes that progressively move away from the deterministic equilibrium, a phenomenon called the \textit{peak drift} \cite{Hanggi1995,Mamis2019d,Mamis2021}. However, the trends in peak drift are opposite in the two perturbations; the distribution mode under white noise moves towards lower numbers of cases, while under Ornstein--Uhlenbeck noise, it shifts towards higher values.

The main difference between the two stochastic perturbations, as shown in Fig. \ref{fig:2}, is the following. Whereas the distribution corresponding to Ornstein--Uhlenbeck noise stays always unimodal, the distribution corresponding to white noise undergoes a bifurcation at $\sigma\simeq0.5\bar{\lambda}$, where an additional mode in the low COVID cases regime ($\simeq 5\times 10^7$ cases) emerges, and the distribution becomes bimodal.
This bimodality enables noise-induces transitions towards the lower number of cases, which in turn explains why the white noise model underestimates the severity of the pandemic (see Fig.~\ref{fig:1}D).

\section{Conclusions}
Our study highlights the challenges of incorporating parameter uncertainties in compartmental epidemiological models. We showed that, although common, modeling these uncertainties with white noise is fundamentally flawed, since the corresponding stochastic models systematically underestimate the number of infected individuals. 

We then proposed a principled modeling of the average contact rate, which accounts for the uncertainties in human social behavior. By considering temporal correlations in each individual's behavior, we showed that the Ornstein--Uhlenbeck process is the correct model for uncertainties in the contact rate. We demonstrated the efficacy of our proposed stochastic model on the SIS model of sexually transmitted disease and the SEIR model of COVID-19 pandemic.

The proposed Ornstein--Uhlenbeck model has one unfavorable characteristic: it allows for rare instances where the contact rate becomes negative, which is not epidemiologically allowed (see Appendix \ref{sec:SEIR}). Nonetheless, the resulting model is amenable to mathematical analysis and produces accurate forecasts that do not suffer from the idiosyncrasies of white noise.
As such, Ornstein--Uhlenbeck process serves as a reliable minimal model of average contact rates.

Finally, the Ornstein--Uhlenbeck process has a time correlation parameter whose value needs to be derived from empirical data; such data is currently very scarce~\cite{Watts2006,Tkachenko2021,mistry2021}. Hence, our work indicates the need for more empirical studies that quantify the correlation time in close proximity interactions.

\section{Materials and Methods}
\subsection{SIS model}
In the SIS model (\ref{eq:SIS}), $N$ denotes the total population. The term $(\lambda/N)SI$ is the simplest form for the disease transmission, and it is based on the assumption of homogeneous mixing of population \cite{Tolles2020a}. Transmission term without the division with $N$ is sometimes used \cite{Gray2011}; however, this choice is not supported by empirical evidence~\cite{Rhodes2008}. Under the constant population assumption $S+I=N$,  the SIS model (\ref{eq:SIS}) can be reduced to one scalar ordinary differential equation. Consider the infected as a fraction of the population by defining $X=I/N$ for the state variable. Then the scalar equation for SIS model reads
\begin{equation}\label{eq:X}
\frac{\id X}{\id t}=\lambda X(1-X)-\gamma X.
\end{equation}
For $R_0>1$, \eqref{eq:X} has an unstable equilibrium at $X= 0$ and a stable equilibrium at $X=(\lambda-\gamma)/\lambda$. The characteristic Lyapunov time of the system is $(\lambda-\gamma)^{-1}$; see Appendix \ref{sec:deterministic_SIS}.  We obtain the stochastic SIS model by perturbing $\lambda$ in \eqref{eq:X} by noise: $\lambda=\bar{\lambda}+\sigma\xi(t)$.  For the case of white noise perturbation, the evolution of distribution of $X$ is governed by the classical Fokker--Planck equation~\cite{Gardiner2004}. Recently, we have also formulated an approximate nonlinear Fokker--Planck equation for the case of correlated noise perturbations~\cite{Mamis2019d,Mamis2021}. Since \eqref{eq:X} is scalar, the stationary solutions of the Fokker--Planck equations are available in closed form in both cases of white noise and correlated noise (see Appendix \ref{sec:stochastic_models}). These closed-form stationary solutions are used in Fig.\ref{fig:3}. For the Monte Carlo simulations, numerical solutions of stochastic SIS and SEIR models are generated by a predictor-corrector scheme~\cite{Cao2015}.

\subsection{SEIR model and fit to COVID data} 
For the SEIR model (\ref{eq:SEIR}), the total population $N$ is assumed constant and equal to the US population of 329.5 million. The data is obtained from the COVID-19 Dashboard by the Center for Systems Science and Engineering (CSSE) at Johns Hopkins University~\cite{JohnsHopkins} and was analyzed using a publicly available MATLAB code~\cite{VargaLajos2020}. December 3, 2021 is considered as the initial time $t_0$, i.e., the start of the Omicron wave in the US.  The parameter values are determined by least squares fitting, resulting in the basic reproduction number $R_0=\lambda/\gamma=1.85$, incubation rate $\alpha=1/3.5$ days$^{-1}$, and curing rate $\gamma=1/1.2$ days$^{-1}$.  The initial values of the exposed $E(t_0)$, the infected $I(t_0)$, and the removed $R(t_0)$ at the beginning of the Omicron wave were chosen consistent with the Johns Hopkins data base to be 0.14\%, 0.18\% and 14.88\% of the total population, respectively. 
To obtain the cumulative COVID-19 cases from the SEIR model,  we use the relation $\int_{t_0}^tI(s)\id s=(R(t)-R(t_0))/\gamma$ from \eqref{eq:removed}, which measures the total number of cases over $[t_0,t]$.

\section*{Acknowledgments}
We are grateful to Prof. Alun Lloyd (North Carolina State University) for fruitful discussions.

\appendix

\section{Deterministic SIS model}\label{sec:deterministic_SIS}
As derived in \eqref{eq:X},  SIS model is expressed as the scalar ordinary differential equation (ODE):
\begin{equation}\label{eq:X2}
\frac{\id X(t)}{\id t}=\lambda X(t)(1-X(t))-\gamma X(t),
\end{equation}
where $X\in[0,1]$ is the number of infected as fraction of the total population, $\lambda$ is the average contact rate, and $\gamma$ the average curing rate, see, e.g., Ref. \cite{Hethcote1984}.  Basic reproduction number is defined as $R_0=\lambda/\gamma$.  
\subsection{Deterministic equilibrium points}\label{sec:deterministic_equil}
The equilibrium points of \eqref{eq:X2} are determined to 0 and $(\lambda-\gamma)/\lambda$.  We also easily derive that:
\begin{itemize}
\item for $R_0<1$, equilibrium point 0 is stable,
\item for $R_0>1$, equilibrium point 0 is unstable and $(\lambda-\gamma)/\lambda$ is stable.
\end{itemize}
Thus, in deterministic SIS models, basic reproduction number $R_0$ determines if the disease dies out ($R_0<1$) or becomes endemic in the population ($R_0>1$). 
\subsection{Characteristic time scale}\label{sec:time_scale}
For $R_0>1$, we derive the characteristic timescale for \eqref{eq:X2}. To this end, we linearize \eqref{eq:X2} around endemic equilibrium $(\lambda-\gamma)/\lambda$, to obtain the equation for variations $\delta X(t)$:
\begin{equation}\label{eq:variation}
\frac{\id}{\id t}\delta X(t)=-(\lambda-\gamma)\delta X(t), \ \ \delta X(t_0)=\delta X_0.
\end{equation} 
Solution of \eqref{eq:X2} is determined to $\delta X(t)=\delta X_0\exp[-(\lambda-\gamma)(t-t_0)]$. Thus, Lyapunov exponent is determined to $\lambda-\gamma$, and thus we identify its inverse $(\lambda-\gamma)^{-1}$ as the characteristic time scale of \eqref{eq:X2}.
\section{Stochastic SIS models}\label{sec:stochastic_models}
Under the stochastic perturbation of contact rate $\lambda=\bar{\lambda}+\sigma\xi(t)$,  where $\bar{\lambda}$ is the mean value and $\sigma$ is the standard deviation of the noise, SIS \eqref{eq:X2} reads
\begin{equation}\label{eq:X_stoch}
\frac{\id X(t)}{\id t}=\bar{\lambda} X(t)(1-X(t))-\gamma X(t)+\sigma X(t)(1-X(t))\xi(t).
\end{equation}
\eqref{eq:X_stoch} is a stochastic differential equation (SDE) under multiplicative noise excitation, since noise $\xi(t)$ is multiplied by a state-dependent function.  For SDE (\ref{eq:X_stoch}), we define the drift and the intensity functions:
\begin{equation}\label{eq:drift_int}
h(x)=\bar{\lambda}x(1-x)-\gamma x, \ \ \sigma(x)=\sigma x(1-x).
\end{equation}
\subsection{SIS model under white noise perturbation}\label{sec:white}
We first study the case where $\xi(t)$ in \eqref{eq:X_stoch} is the Gaussian standard white noise with zero mean value and autocorrelation $\mathbb{E}[\xi(t)\xi(s)]=\delta(t-s)$, where $\delta(t-s)$ is Dirac's delta function. Under It\=o interpretation of stochastic differential \eqref{eq:X_stoch}, see Ref. \cite[Sec. 5.4.3]{Sun2006}, the probability density function (PDF) $p(x,t)$ of $X(t)$ is governed by the classical Fokker-Planck equation (see, e.g., Ref.\cite{Gardiner2004}):
\begin{equation}\label{eq:FP}
\frac{\partial p(x,t)}{\partial t}+\frac{\partial}{\partial x}[h(x)p(x,t)]=\frac{1}{2}\frac{\partial^2}{\partial x^2}\left[\sigma^2(x)p(x,t)\right].
\end{equation} 
The closed-form stationary solution $p_0(x)=\lim_{t\rightarrow\infty}p(x,t)$ of the Fokker-Planck \eqref{eq:FP} is given by \cite[Sec. 5.3.3]{Gardiner2004}
\begin{equation}\label{eq:p0}
p_0(x)=\frac{\mathcal{N}}{\sigma^2(x)}\exp\left(2\int^x\frac{h(y)}{\sigma^2(y)}\id y\right),
\end{equation}
where $\int^x\id y$ denotes the antiderivative, and $\mathcal{N}$ is the normalization factor so that $\int_{\mathbb{R}}p_0(x)\id x=1$.  Specification of \eqref{eq:p0} under \eqref{eq:drift_int} results in
\begin{equation}\label{eq:p0_white}
p_0(x)=\mathcal{N}x^{2\left(1-V-R_0^{-1}\right)/V}(1-x)^{-2\left(1+V-R_0^{-1}\right)/V}\exp\left(-\frac{2}{VR_0}\frac{1}{1-x}\right).
\end{equation}
The two dimensionless parameters that appear in \eqref{eq:p0_white} are the basic reproduction number of the original deterministic model, $R_0=\bar{\lambda}/\gamma$, and the relative variance $V=\sigma^2/\bar{\lambda}$ of white noise. Having $p_0(x)$ in closed form (\ref{eq:p0_white}) we are able to examine the PDF form. The result we obtained after algebraic calculations, are summarized below:
\begin{itemize}
\item In the vicinity of 0, $p_0(x)\sim x^{2\left(1-V-R_0^{-1}\right)/V}$. For
\begin{equation}\label{eq:cond1}
\frac{2}{V}\left(1-V-R_0^{-1}\right)<-1\Rightarrow V>2(1-R_0^{-1}),
\end{equation}
$p_0(x)$ of \eqref{eq:p0_white} is not integrable, and $\lim_{x\rightarrow 0^+}p_0(x)=+\infty$. Thus, under \eqref{eq:cond1}, $p_0(x)$ is a delta function at 0.  Note that \eqref{eq:cond1} is expressed equivalently as $R_0-\sigma^2/(2\gamma)<1$, which is the disease extinction condition stated in Ref. \cite{Gray2011}.  \eqref{eq:cond1}  always holds true for $R_0<1$, resulting thus in disease eradication, as in the deterministic case. This is the reason for considering $R_0\in[1,+\infty]$ in our analysis. Furthermore, \eqref{eq:cond1} illustrates the unrealistic result discussed in the main article,  that increase in white noise intensity makes disease extinction more likely.   
\item For $V<2(1-R_0^{-1})$ of \eqref{eq:p0_white} is integrable.  Additionally, for
\begin{equation}\label{eq:cond4}
\frac{2}{V}\left(1-V-R_0^{-1}\right)>0\Rightarrow V<1-R_0^{-1}, 
\end{equation}
$p_0(x)$ is unimodal, exhibiting its maximum at a nonzero $x$.
\item On the other hand, for
\begin{equation}\label{eq:cond2}
-1<\frac{2}{V}\left(1-V-R_0^{-1}\right)<0\Rightarrow 1-R_0^{-1}<V<2(1-R_0^{-1}),
\end{equation}
$p_0(x)$ exhibits a maximum at 0.  Under the simultaneous satisfaction of \eqref{eq:cond2} and
\begin{equation}\label{eq:cond3}
R_0>\frac{8V}{(V+1)^2},
\end{equation}
$p_0(x)$ is bimodal, with one mode at 0. Otherwise, $p_0(x)$ is unimodal at 0.
\end{itemize}
By using the above results, we formulate the bifurcation diagram in Fig. \ref{fig:3}.

\subsection{SIS model under Ornstein--Uhlenbeck perturbation}\label{sec:OU}
Now, we move on to the case where $\xi(t)$ in \eqref{eq:X_stoch} is a standard Ornstein--Uhlenbeck process, with zero mean value and autocorrelation 
\begin{equation}\label{eq:OU_auto}
\mathbb{E}[\xi(t)\xi(s)]=\frac{1}{2\tau}\exp\left(-\frac{|t-s|}{\tau}\right),
\end{equation}
with $\tau$ being the correlation time. For this case, in our recent work \cite{Mamis2021}, we derived an approximated nonlinear Fokker-Planck equation for $p(x,t)$:
\begin{equation}\label{eq:nFP}
\frac{\partial p(x,t)}{\partial t}+\frac{\partial}{\partial x}\left\{\left[h(x)+\sigma'(x)\sigma(x)A(x,t;p)\right]p(x,t)\right\}=\frac{\partial^2}{\partial x^2}\left[\sigma^2(x)A(x,t;p)p(x,t)\right],
\end{equation}
where coefficient $A(x,t;p)$ is defined as
\begin{equation}\label{eq:A}
A(x,t;p)=\sum_{m=0}^2\frac{D_m(t;p)}{m!}\left\{\zeta(x)-\mathbb{E}\left[\zeta\left(X(t)\right)\right]\right\}^m,
\end{equation}
where
\begin{equation}\label{eq:zeta}
\zeta(x)=\sigma(x)\left(\frac{h(x)}{\sigma(x)}\right)'=-\gamma \frac{x}{1-x},
\end{equation}
and
\begin{equation}\label{eq:D}
D_m(t;p)=\frac{1}{2\tau}\int_{t_0}^t\exp\left(-\frac{t-s}{\tau}\right)\exp\left(\int_{s}^t\mathbb{E}\left[\zeta(X(u))\right]\id u\right)(t-s)^m\id s.
\end{equation}
Fokker-Planck equation (\ref{eq:nFP}) is nonlinear, due to the dependence of $A(x,t;p)$ on the moment $\mathbb{E}[\zeta(X(t))]$. Its stationary solution reads
\begin{equation}\label{eq:p0_OU}
p_0(x,R)=\frac{\mathcal{N}}{\sigma(x)A(x,R)}\exp\left(\int^x\frac{h(y)}{\sigma^2(y)A(y,R)}\id y\right),
\end{equation}
where $A(x,R)$ is the asymptotic value of coefficient $A(x,t;p)$:
\begin{equation}\label{eq:A0}
A(x,R)=\frac{1}{2}\sum_{m=0}^2\frac{[\tau(\zeta(x)-R)]^m}{(1-\tau R)^{m+1}},
\end{equation}
and $R$ is the asymptotic value of moment $\mathbb{E}[\zeta(X(t))]$.  As presented in Ref.  \cite{Mamis2021}, the unknown $R$ in closed-form solution (\ref{eq:p0_OU}) is determined by using an iteration scheme, based on the definition of moment $R$:
\begin{equation}\label{eq:R}
R=\int_{\mathbb{R}}\zeta(x)p_0(x,R)\id x.
\end{equation}
The distributions of the asymptotic infected fraction under Ornstein--Uhlenbeck perturbation, that we plot in Fig. \ref{fig:3}, are determined by using \eqref{eq:p0_OU} and the said iteration scheme. Note that they are in excellent agreement with the distribution obtained from direct Monte Carlo simulations, as shown in Fig. \ref{fig:3}.


\begin{figure}
	\centering
	\includegraphics[width=\textwidth]{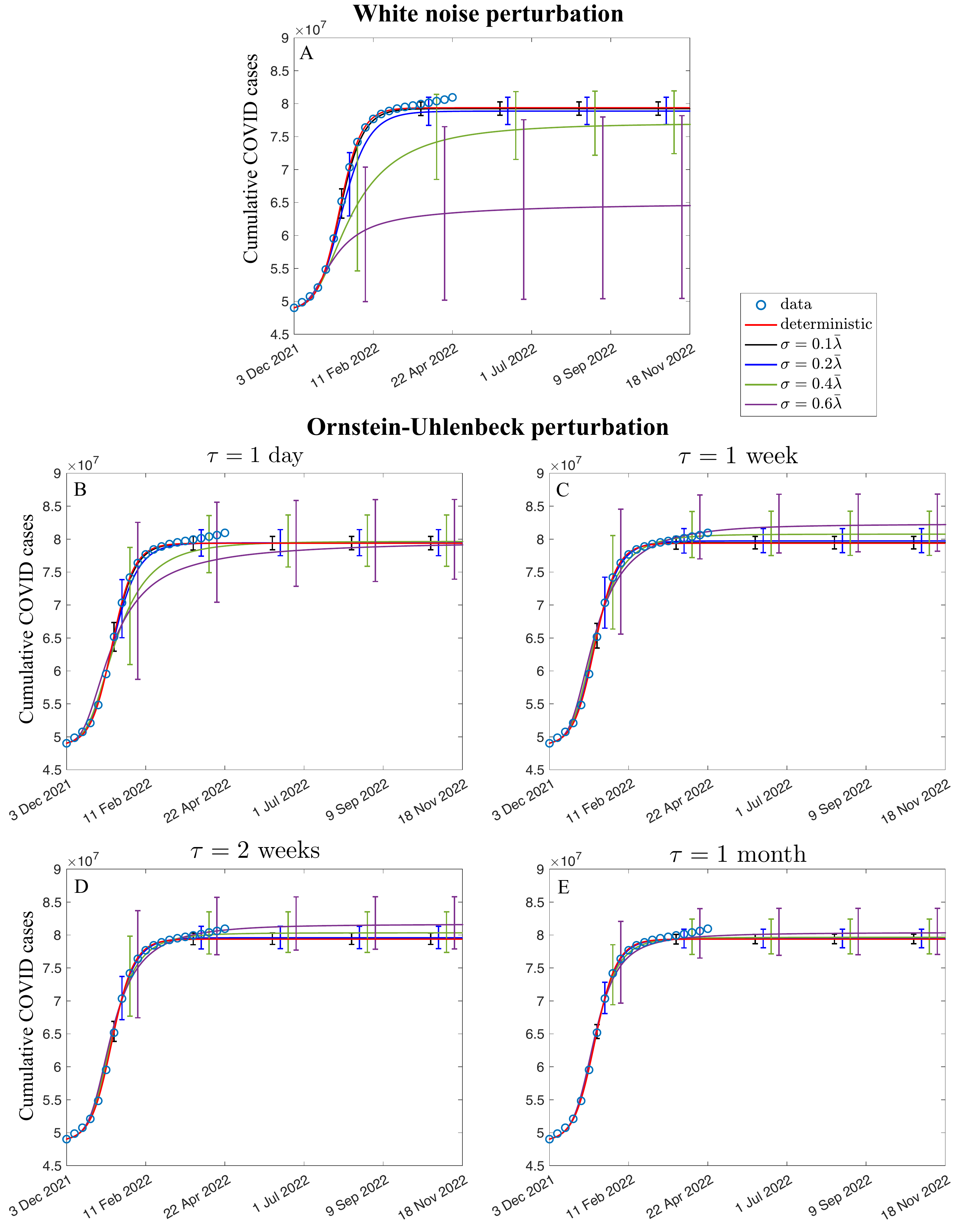}
	\caption{Predictions (mean trajectories with 50\% errorbars) of stochastic SEIR models for the cumulative number of COVID cases in the US during omicron wave, for increasing noise intensity.  \textbf{A:} white noise perturbation of contact rate $\bar{\lambda}$. \textbf{B-E:} Ornstein-Uhlenbeck perturbations of $\bar{\lambda}$ with correlation time ranging from 1 day to 1 month.}\label{fig:SI1}
\end{figure}

\section{Stochastic SEIR models}\label{sec:SEIR}
SEIR model (\ref{eq:SEIR}), after stochastically perturbing its contact rate; $\lambda=\bar{\lambda}+\sigma\xi(t)$ reads:
\begin{subequations}\label{eq:stoch_SEIR}
\begin{equation}
\frac{\id S}{\id t}=-\frac{\bar{\lambda}}{N}SI+\frac{\sigma}{N}SI\xi(t),
\end{equation}
\begin{equation}
\frac{\id E}{\id t}=\frac{\bar{\lambda}}{N}SI-\alpha E+\frac{\sigma}{N}SI\xi(t),
\end{equation}
\begin{equation}
\frac{\id I}{\id t}=\alpha E-\gamma I,
\end{equation}
\begin{equation}
\frac{\id R}{\id t}=\gamma I.
\end{equation}
\end{subequations}
For stochastic SEIR model (\ref{eq:stoch_SEIR}), we generated 50000 trajectories using the predictor-corrector scheme proposed in Ref. \cite{Cao2015}. Since this scheme works for SDEs under white noise excitation, it was applied directly to SDE system (\ref{eq:stoch_SEIR}) in the case of perturbation being white noise; $\xi(t):=\xi^{\text{WN}}(t)$. For $\xi(t)$ being a standard Ornstein--Uhlenbeck process, SDE system (\ref{eq:stoch_SEIR}) was augmented by the equation
\begin{equation}\label{eq:OU_SDE}
\frac{\id \xi(t)}{\id t}=-\frac{1}{\tau}\xi(t)+\frac{1}{\tau}\xi^{\text{WN}}(t),
\end{equation}
which is the SDE that has white noise as input, and generates an Ornstein--Uhlenbeck process with zero mean and autocorrelation given by \eqref{eq:OU_auto}, see Refs. \cite{Hanggi1995,Gardiner2004}.

\subsection{Effect of correlation time}\label{sec:corr_time}
In Fig.  \ref{fig:SI1}, we present a more detailed survey of the effect of correlation time $\tau$ of contact rate perturbation $\xi(t)$.  Figs.  \ref{fig:SI1}A,C are also shown in Fig. \ref{fig:1}C of the main paper,  depicting the mean trajectories with 50\% errorbars for the COVID cases, under white noise perturbation (Fig. \ref{fig:SI1}A), and Ornstein-Uhlenbeck perturbation with correlation time $\tau=1$ week.  

In Fig.\ref{fig:SI1} we observe that the mean trajectory of SEIR model under Ornstein--Uhlenbeck perturbation stays always close to the deterministic prediction, for all choices of $\tau$ we considered.  For the shortest correlation time of 1 day, see Fig.\ref{fig:SI1}B, the mean asymptotic number of COVID cases coincides with the prediction of the deterministic model, for all noise intensities we considered. The only effect we observe by increasing noise intensity is that the asymptotic number of cases is attained after a longer time period. 

For longer correlation times of 1 week to 1 month (see Figs.\ref{fig:SI1}C-E), increase in noise intensity results in a slight increase in the asymptotic number of cases predicted in the mean by stochastic SEIR model.  However, this drift towards larger numbers of cases is decreased by increasing the correlation time: For noise intensity being 60\% of the average contact rate, the mean prediction of asymptotic number of cases increases by 3.8\% for $\tau=1$ week, by 2.8\% for $\tau=2$ weeks, and by  1.2\% for $\tau=1$ month. 

\begin{figure}
	\centering
	\includegraphics[width=\textwidth]{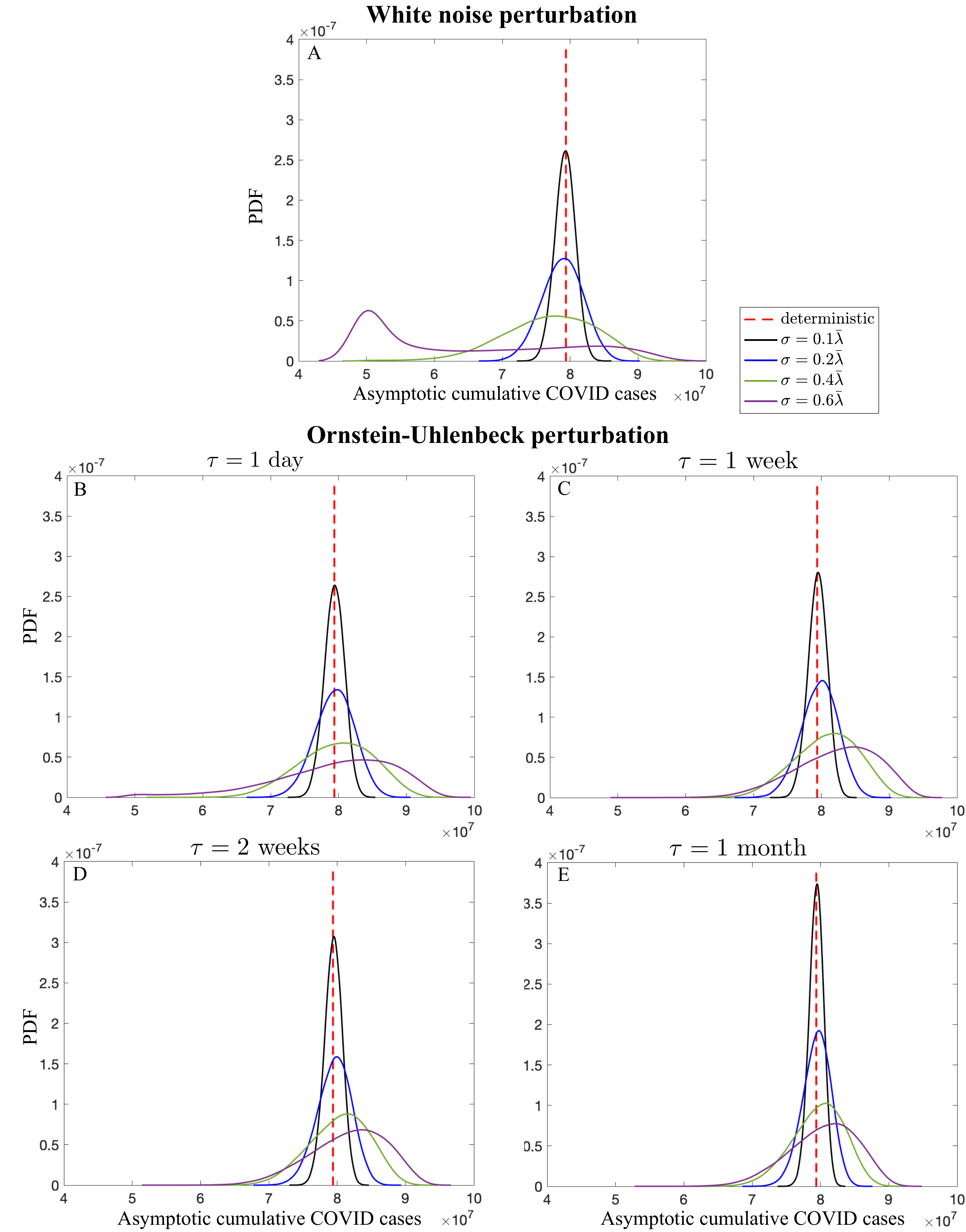}
	\caption{PDF of the asymptotic cumulative number of COVID cases in the US during omicron wave, for increasing noise intensity.  \textbf{A:} SEIR with white noise perturbation of contact rate $\bar{\lambda}$. \textbf{B-E:} Ornstein--Uhlenbeck perturbations of $\bar{\lambda}$ with correlation time ranging from 1 day to 1 month.}\label{fig:SI2}
\end{figure}
\thispagestyle{empty}

In Fig.\ref{fig:SI2}, we depict the PDFs for the asymptotic number of cases, as calculated from the stochastic SEIR models. As we have commented in the main article, the peak drift phenomenon (see Refs. \cite{Hanggi1995,Mamis2019d,Mamis2021}) has opposite trends in the PDFs under the two perturbations. For the PDF under white noise, and by increasing noise intensity, the PDF mode shifts from the deterministic equilibrium towards lower number of cases. Increase in the intensity of Ornstein-Uhlenbeck perturbation results in the shift of PDF mode towards higher number of cases. In the  main article, we have also observed that, for $\sigma=0.6\bar{\lambda}$, an additional mode around $5\cdot10^7$ cases has emerged in the PDF under white noise. This noise-induced transition in the PDF shape is suppressed if we consider correlated noise.  However, an insignificant peak around $5\cdot10^7$ cases is also detected for $\sigma=0.6\bar{\lambda}$, for short correlation time $\tau=1$ day. For larger correlation times, from 1 week to 1 month, the PDF always stays unimodal. Last, increase in correlation time results in the PDF to be less diffusive,  concentrated around the deterministic equilibrium. This is the sharpening effect of the correlated noise, see Refs. \cite{Hanggi1995,Mamis2021}, and is also observed in Fig.\ref{fig:SI1}B-E, where increase in $\tau$ results in shrinking errorbars around the mean trajectories.  

\subsection{On the perturbed contact rate attaining negative values}\label{sec:neg}
\begin{figure}
	\centering
	\includegraphics[width=0.8\textwidth]{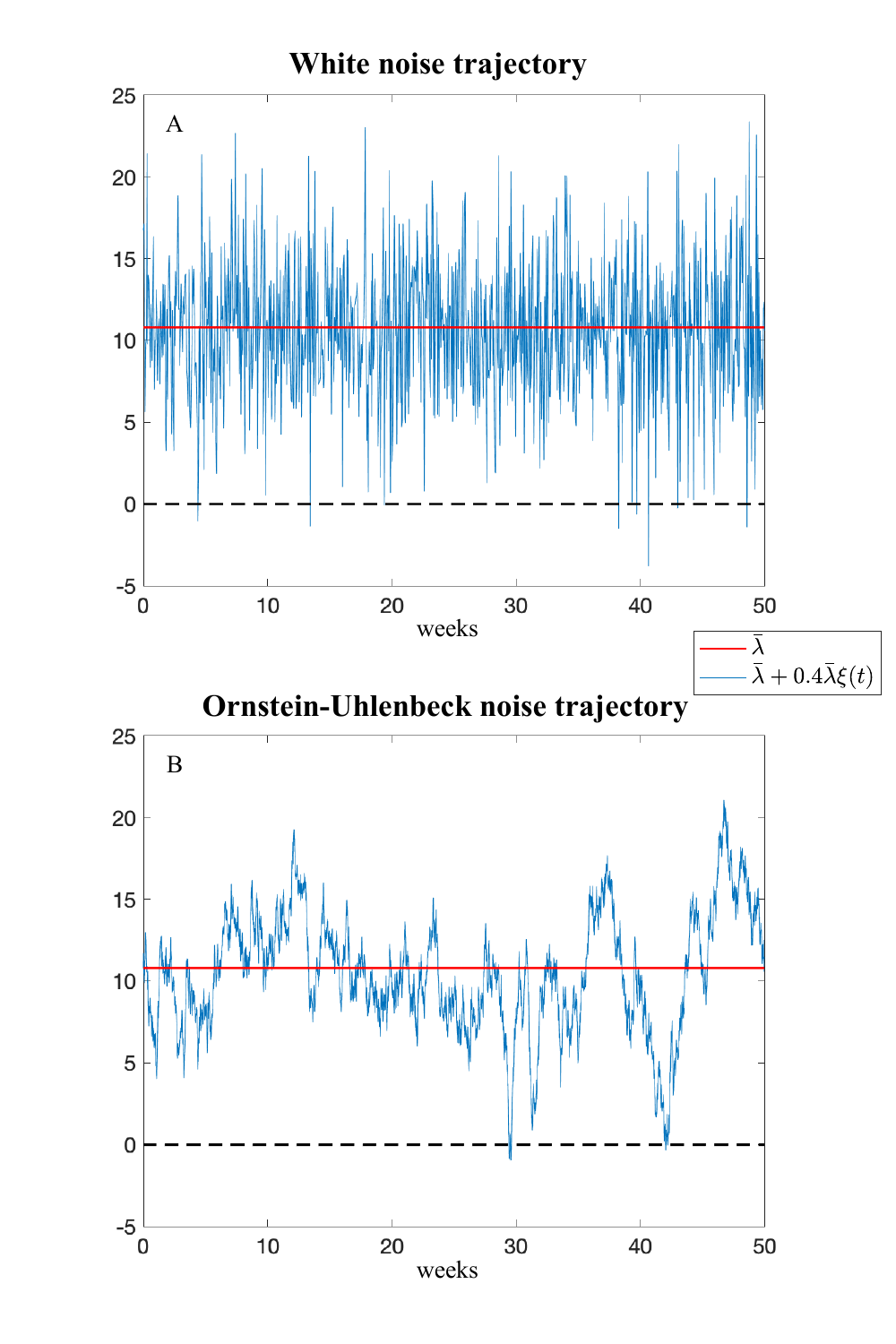}
	\caption{Trajectory of the contact rate for COVID perturbed around its mean value $\bar{\lambda}= 10.8$ weeks$^{-1}$ (shown by the red line) by noise of intensity $\sigma=0.4\bar{\lambda}$. Black dashed line indicates the zero level. \textbf{A:} white noise perturbation. \textbf{B:} Ornstein--Uhlenbeck noise perturbation, with correlation time $\tau=1$ week.}\label{fig:SI3}
\end{figure}

In Ref. \cite{DOnofrio2008}, a behavior similar to that of stochastic compartmental models was observed. Incorporation of Gaussian white noise in an ODE modeling tumor growth, resulted in the eradication of the tumor. In the aforementioned work, this result was attributed to the unboundedness of Gaussian noise: stochastic perturbation by Gaussian noise means that positive parameters (like tumor growth or contact rate) can also attain negative values, which would violate their physical meaning.  As we observe in Fig.\ref{fig:SI3}B, the Ornstein--Uhlenbeck perturbation, despite also being Gaussian, results in far less negative values of the perturbed $\lambda$, compared to white noise perturbation (see Fig. \ref{fig:SI2}A). This is due to the mean-reverting property of Ornstein--Uhlenbeck process.


\end{document}